\documentclass[12pt]{iopart}
\usepackage{iopams}
\usepackage{graphicx} 
\usepackage{float}
\expandafter\let\csname equation*\endcsname\relax
\expandafter\let\csname endequation*\endcsname\relax
\usepackage{amsmath}
\usepackage{amsfonts}
\usepackage{amssymb}
\usepackage[spaces,hyphens]{url}
\usepackage{hyperref}
\usepackage{caption}
\usepackage{subcaption} 
\usepackage[numbers,sort&compress]{natbib}
\graphicspath{{gfx/}}
\begin{document}

\title[Universality class of synchronization transition with disorder]{Synchronization transition in space-time chaos in the presence of quenched disorder}

\author{Naval R. Sabe$^a$, Priyanka D. Bhoyar$^b$, and Prashant M. Gade$^a$}

\address{(a) Department of Physics, Rashtrasant Tukadoji Maharaj Nagpur University,
campus, Nagpur, 440033, India. (b) Department of Physics, Seth Kesarimal Porwal College of Arts and Science and Commerce, Kamptee, Nagpur, 441001, India.}
\ead{prashant.m.gade@gmail.com}
\vspace{10pt}

\begin{abstract}
Synchronization of two replicas of coupled map lattices for continuous maps is known to be in the multiplicative noise universality class.
We study this transition in the presence of quenched disorder in coupling. The
disorder is identical in both replicas. We study one-dimensional, two-dimensional, and globally coupled logistic and tent maps.  We observe a clear second-order transition with new exponents. The order parameter decays as $t^{-\delta}$ and $\delta$ depends on the map
and its parameters. 
The asymptotic order parameter for $\Delta$ distance from a critical point
grows as $\Delta^{\beta}$ with $\beta=\delta$.
The quenched disorder in coupling is a relevant perturbation for the replica synchronization of coupled map lattices.
The critical exponents are different from those of
the multiplicative noise universality class. However, it does not depend on dimensionality if the transition is continuous for the cases studied.
\end{abstract}
\section{Introduction:}
Synchronization is everywhere in nature: neuronal populations,
cardiac pacemakers, Josephson circuits, power-grid networks,
lasers, and even coupled chaotic systems can synchronize during their activity.
We have developed some understanding of these phenomena using the
nonlinear dynamics framework~\cite{pikovsky2001universal}. Synchronization can be viewed as an absorbing phase transition.  Once the system is synchronized, it cannot come out of the synchronized state. The detailed balance condition is not satisfied. This is
a nonequilibrium phase transition. We can study it from
the phase transition viewpoint. 
Necessary conditions for synchronization of
spatiotemporal systems even for in thermodynamic limit
are well-understood now~\cite{PhysRevLett.64.821,gade1996synchronization}
from the viewpoint of linear stability analysis.
  There are very few studies of this type.
For absorbing phase transitions in stochastic systems, few universality classes have been proposed.
The most popular and widely observed universality class is
directed percolation (DP). Few other universality classes such as
compact directed percolation, Manna class,
dynamical percolation, and voter model have been obtained~\cite{henkel2008non}. 
Such studies are also extended to spatially extended chaotic systems.
For example, directed percolation transition has been observed in coupled
circle maps~\cite{janaki2003evidence}. Of course, we do not have an analog of 
all different transitions in nonequilibrium statistical mechanics in coupled map
lattices.  Coupled map lattices (CML)~\cite{kaneko1993theory},
which are prototype models for systems displaying spatiotemporal chaos.
CML can be seen as time and space discrete versions of partial
differential equations. Studies in CML from the viewpoint of phase transition are particularly
fascinating because they connect the nonlinear dynamics of spatiotemporal
systems with nonequilibrium statistical physics.

A transition in the universality class of bounded Kardar-Parisi-Zhang (BKPZ) universality
class was found in the synchronization between two spatially extended systems 
by Ahlers and Pikovsky.
They showed that the transition for one-dimensional systems of coupled maps
is generically in the BKPZ class for continuous maps ~\cite{PhysRevLett.88.254101}. 
It has now been established that if the coefficient of the KPZ non-linearity is positive, the
introduction of an upper wall leads to a  set of critical exponents characterizing the multiplicative noise 1 (MN1) universality class~\cite{grinstein1996phase,tu1997systems,munoz1998nonlinear,munoz2004multiplicative,genovese1999recent}.
On the contrary, the same transition is in the universality class of DP
for discontinuous maps~\cite{PhysRevLett.88.254101}.
There has been no experimental
evidence of the MN class to date, whereas DP critical exponents have
recently been measured experimentally in one~\cite{rupp2003critical}
and two spatial dimensions~\cite{takeuchi2007directed}. The disorder is
inevitable in experimental systems. Mu{\~n}oz has argued that "The presence of quenched
disorder is known to generate new non-trivial phenomenology in both Ising and DP-like
systems. It is also
expected to be a relevant perturbation at the different MN classes reported here."\cite{munoz2004nonequilibrium}.  Hence it is important to study the
effect of quenched disorder. 

The effect of quenched disorder in equilibrium systems was studied by Griffiths
who showed that \textquotedblleft in a class of randomly diluted Ising ferromagnets
the magnetization fails to be an analytic function of the field 
$H$ for $H=0$ for a range of temperatures above that at which spontaneous magnetization
first appears\textquotedblright~\cite{griffiths}. Later, the effect of quenched disorder in nonequilibrium systems was studied in
great detail~\cite{vojta}. In this case, the necessary condition for the quenched disorder 
to be a relevant disorder is known as the Harris criterion. It states that for $d\nu<2$ 
where $\nu$ is the correlation length exponent, the disorder is relevant.   An 
interesting manifestation of this possibility is observed in the contact process.
We observe a transition in 
activated scaling universality class where power-law
decay of the order parameter with continuously changing exponents
is obtained in a range of parameters, This is known as Griffiths phase.
 There have been experimental investigations in critical
behavior of disordered antiferromagnets\cite{belanger2000experimental}

Thus there are cases where quenched disorder is a relevant 
perturbation. It could be a relevant perturbation for the MN1 class.
 However, the effect of quenched disorder on systems
in the MN universality class is not studied to
the best of our knowledge. In this work, we study this effect 
in the context of the synchronization of replicas of coupled map
lattices.
We find that if the transition is in 
DP universality class, the quenched disorder leads to Griffiths 
phase while we obtain a new universality class if this
transition is in MN class. 

It has been argued that the synchronization transition
belongs to the MN or DP universality class, depending on whether
the system has linear or nonlinear instabilities~\cite{cencini2001linear}. 
Both classes are obtained for identical one-dimensional
CML replicas~\cite{ahlers2002critical}. These studies are extended to
two dimensions~\cite{ginelli2009synchronization}. Even for 
the synchronization of two delayed chaotic systems, which
can be modeled as a synchronization of two 1-d spatiotemporal systems, MN as well
as DP universality class is observed~\cite{szendro2005universal}.  It has been shown that
instead of two different maps, a single map
can show both DP and MN universality class for different parameter values~\cite{bagnoli2006synchronization}. Similarly,  both MN and DP 
universality classes are obtained on varying control parameters in wetting transition~\cite{ginelli2003multiplicative}. Replica synchronization is also studied in
the presence of long-range spatial interactions where continuously changing exponents are observed~\cite{10.1063/1.2945903}. The stochastic theory of these transitions has been 
proposed in~\cite{PhysRevLett.90.204101}. 

Investigation into the effect of quenched disorder is important from a theoretical
viewpoint as mentioned above. Because the disorder is unavoidable in 
real-life systems, it has practical significance as well.
The presence of a mismatch is analogous to the influence of an external field in this paradigm, and hence may still be
accounted for in the framework of the nonequilibrium phase
transitions~\cite{ginelli2009synchronization}.
\section{Model and Simulation} 
Ahlers and Pikovsky studied synchronization transition
in replicas of smooth Coupled Map lattice~\cite{PhysRevLett.88.254101}.
The model is as follows:
\begin{equation*}
	\begin{pmatrix}
		u_1(x,t+1)\\
		u_2(x,t+1)
	\end{pmatrix}
	=
	\begin{pmatrix}
		1-\gamma & \gamma \\
		\gamma & 1-\gamma
	\end{pmatrix}
	\begin{pmatrix}
		(1-2\epsilon(i))f_1(x,t)+\epsilon(i)[f_1(x+1,t)+f_1(x-1,t)] \\
		(1-2\epsilon(i))f_2(x,t)+\epsilon(i)[f_2(x+1,t)+f_2(x-1,t)]
	\end{pmatrix}
\end{equation*}
Here, x=1,2,3$\dots$L and t=1,2,3$\dots$ are the discrete
space and time variables and $\epsilon(i)$ represents the coupling within each CML
whereas $\gamma$ represents the strength of site-wise interaction between
two CML.
$f(u_1)$ or $f(u_2)$ is a nonlinear function that
describes the local dynamics.
The synchronization is studied for $\epsilon(i)=\epsilon=1/3$
 varying values of coupling parameter $\gamma$.
The synchronization is achieved when a synchronization error
$w(x,t)=u_1(x,t)-u_2(x,t)=0$ for all x at
$t \rightarrow \infty$. 
The order parameter is defined as 
$\langle|w(x,t)|\rangle$=$\frac{1}{N}\sum_{i=1}^N |w(x,t)|$ which
is zero in synchronized state. In unsynchronized state $|w(x,t)|>0$. The synchronized
state is an absorbing state and the system cannot come out of it.

\begin{figure}[ht]
	\centering 
	\includegraphics[width=.45\linewidth]{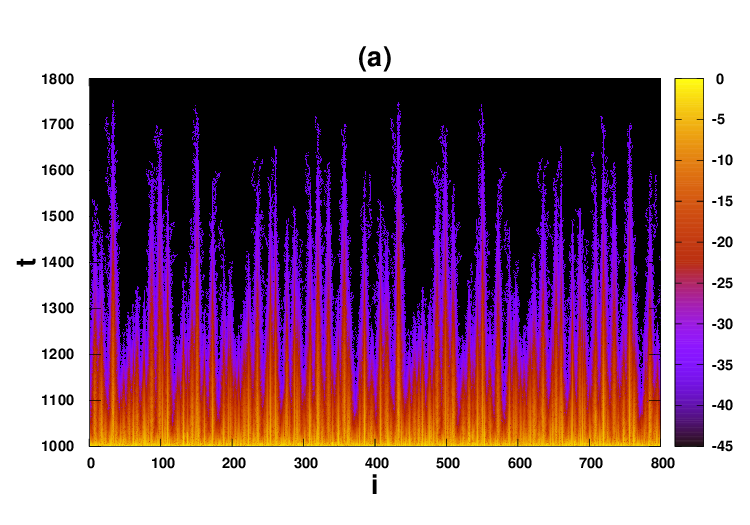} 
	\includegraphics[width=.45\linewidth]{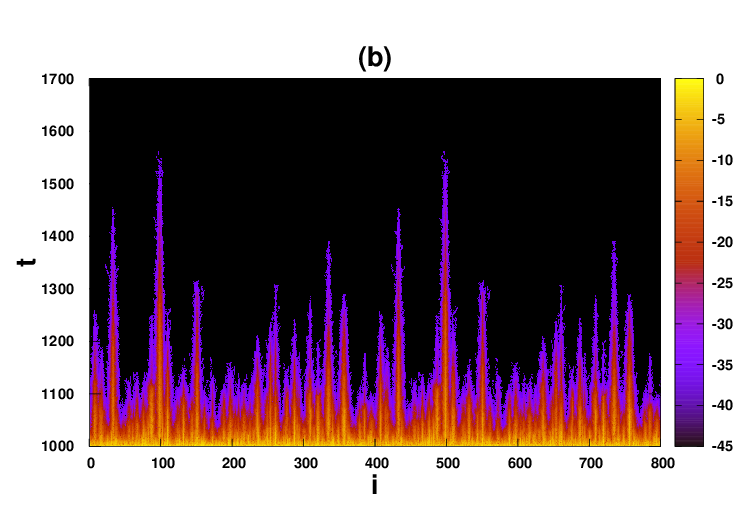}
	\caption{The spatiotemporal evolution of order parameter $\ln\langle|w(x,t)|\rangle$ for  $L=800$ is shown in figure
		at $\gamma$ slightly greater than $\gamma_c$.
		(a) tent CMl with $\epsilon$ disorder at $\gamma=0.255$ in 1D.  
		(b) logistic CML with $\epsilon$ disorder at $\gamma=0.251$ in 1D.
		In both cases, the coupling is switched on after t=1000.
		The value decreases from yellow to black.}
	\label{fig1}  
\end{figure}

They found that the synchronization transition of CML consisting of continuous maps such as tent maps and logistic maps
belongs to the MN universality class. The MN universality class is characterized by
characteristic exponents $\delta=1.1$, $\beta=1.5$, $z=1.53$~\cite{PhysRevLett.88.254101}.  
For discontinuous maps such as the Bernoulli map the transition belongs to
DP universality class and is characterized by critical
exponent $\delta=0.158$, $\beta=0.27$, $z=0.276$~\cite{henkel2008non}.

We find that under the effect of quenched disorder, the model exhibits novel universality class
with critical coefficients different than that of MN universality 
class for continuous maps such as tent and logistic maps.
We study the synchronization transition with a quenched disorder in coupling
strength. 
We study two replicas of a 1D coupled map
lattice of size $L=5\times 10^5$ for a long time $10^6$. We average over $500$
initial conditions and disorder configurations.
We use the tent map: $f(x)=ax \;\; {\rm{if}} \;\; x<1/2 {\rm{and}} \;\; f(x)=a(1-x) {\rm{if}} \;\; x\ge 1/2$ where $a=2$. We also study the
 logistic map: $\mu x(1-x)$ where $\mu =4$ and $0\le x\le 1$.  
(For quenched disorder in parameter $a$ or $\mu$, the synchronization
transition  remains
in the MN universality class.)
We find that the quenched disorder in the coupling parameter $\epsilon(i)=\sigma$,
where $\sigma$ is a uniform
a random number between $(0,0.5)$ is a relevant
perturbation and critical behavior
is altered for synchronization transition. These are maps on an interval
and with this prescription, none of the coupling will be negative avoiding 
possible numerical instability.

\begin{figure}[ht]
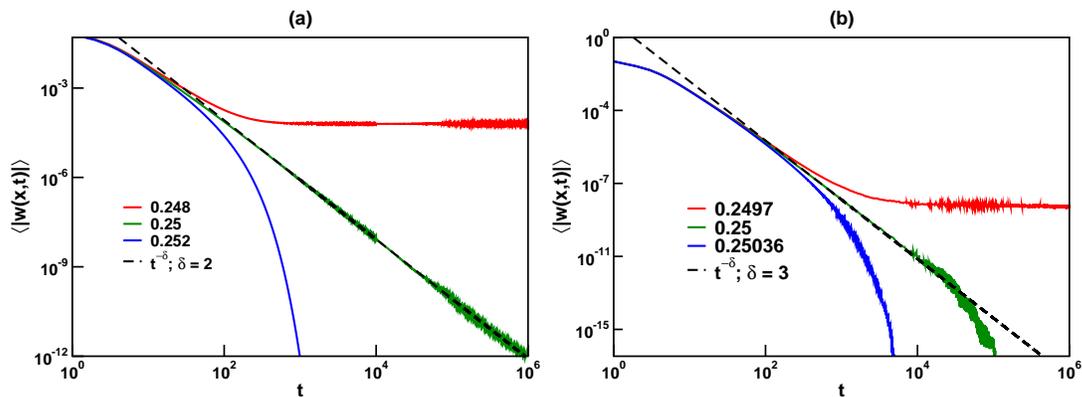

	\centering 
	\includegraphics[width=.45\linewidth]{1dtentop} 
	\includegraphics[width=.45\linewidth]{1dlogop}
	\caption{(a) shows plot of $\langle|w(x,t)|\rangle $ with time for coupled tent 1D CML with a disorder in diffusion parameter
		$\epsilon$ for various values of $\gamma$=0.248,0.25,0.252(top to bottom). The order parameter decays as
		$\langle|w(x,t)|\rangle$ $\sim t^{-\delta}$ where $\delta=2$ at $\gamma_c=0.25$. 
		(b) shows plot of $\langle|w(x,t)|\rangle$ with time at $\gamma=0.2497,0.25,0.25036$ (top to bottom)
		for coupled logistic 1D CML $\epsilon$ disorder. 
		The order parameter decays as $\langle|w(x,t)|\rangle$ $\sim t^{-\delta}$ where $\delta=3$ at $\gamma=0.25$.} 
	\label{fig2}
\end{figure}

The spatiotemporal evolution of synchronization error $\langle|w(x,\infty)|\rangle$ 
is plotted for $L=800$ and $\gamma>\gamma_c$ (see Fig.\ref{fig1}).
In both cases, the disorder coupling is switched on after $t=1000$ transients.
We obtain power-law decay at $\gamma_c=0.25$, $\langle|w(x,t)|\rangle \sim t^{-\delta}$.
Decay coefficient $\delta=2$ and $\delta =3$ for tent map and
logistic map, respectively.(see Fig.\ref{fig2}).
(For MN universality class $\delta=1.1$).
For $\gamma<\gamma_c$, $\langle|w(x,\infty)|\rangle >0$.
For $\gamma>\gamma_c$, $\langle|w(x,t)|\rangle \rightarrow 0$. 

\begin{figure}[ht]
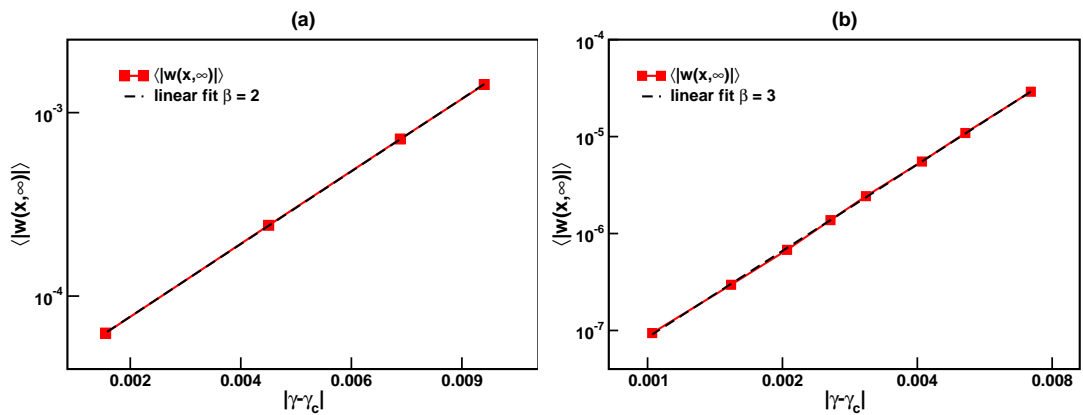
 
	\centering
	\includegraphics[width=.45\linewidth]{1dtentbeta} 
	\includegraphics[width=.45\linewidth]{1dlogbeta} 
	\caption{(a) Shows the plot of $\langle|w(x,\infty)|\rangle$ for several values of $\gamma<\gamma_c=0.25$ in the range 0.24-0.248 tent map. The obtained value of $\beta=2$ in 1D.
		(b) Shows the plot of order parameter $\langle|w(x,\infty)|\rangle$ for several values of 
		$\gamma<\gamma_c=0.25$ in the range 0.248-0.2497 for the logistic map. The obtained value of $\beta=3$ in 1D.} 
	\label{fig3} 
\end{figure}

\begin{figure}[ht]
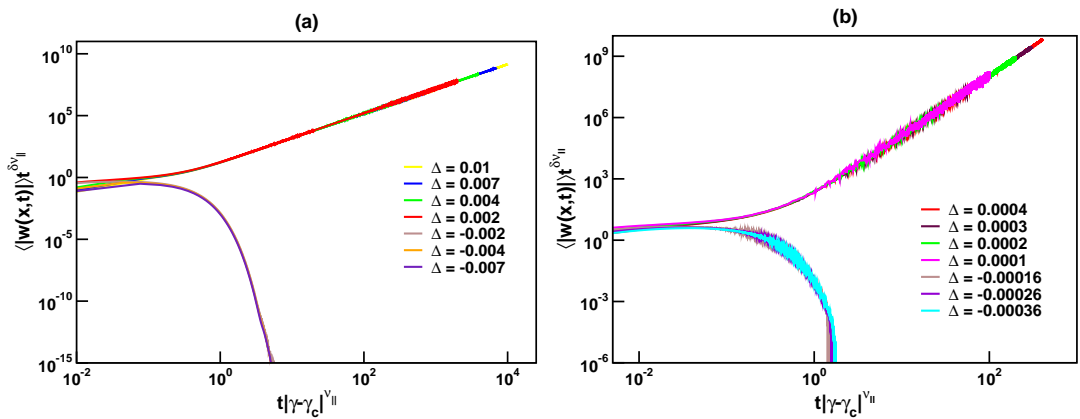

	\centering    
	\includegraphics[width=.45\linewidth]{1dtentoffcr} 
	\includegraphics[width=.45\linewidth]{1dlogoffcri} 
	\caption{(a) Shows the off-critical scaling of the order parameter $\langle|w(x,t)|\rangle$ for
		$-0.01\le \gamma-\gamma_c\le 0.05$ in case of 1D tent CML. 
		(b)Shows the off-critical scaling of the order parameter $\langle|w(x,t)|\rangle$ for 
		$-0.00036\le \gamma-\gamma_c\le 0.0004$ in case of 1D logistic CML.
		The best collapse is obtained for $\nu_{\parallel}=1$ in both cases.} 
	\label{fig4}
\end{figure}

For $\gamma > \gamma_c$, $\langle|w(x,\infty|\rangle>0$. The asymptotic order 
parameter obey the scaling relation $\langle|w(x,\infty|\rangle \sim |\gamma-\gamma_c|^{\beta}$.
The characteristic exponent $\beta=2$ in case of tent map and $\beta=3$
in the case of a logistic map is shown in Fig.\ref{fig3}.
The obtained values of $\beta$ for the tent and logistic map do not match 
that for the MN class.
The off-critical scaling of order parameter $\langle|w(x,t)|\rangle$
is used to obtain the critical exponent $\nu_\parallel$
using the following scaling relation:
$\langle|w(x,t)|\rangle$ $\sim t^{\delta \nu_\parallel} |\gamma-\gamma_c|^{\nu_\parallel}$
for several
values of $\gamma<\gamma_c$ and $\gamma>\gamma_c$.
The off-critical scaling for the tent map and logistic map is shown in Fig.\ref{fig4}.
The obtained value of critical parameters is
different from that of the MN universality class.
The characteristic exponents $\delta$, $\beta$ and $\nu_\parallel$ are
obtained using the hyperscaling relation $\delta = \frac{\beta}{\nu_\parallel}$ in 
both cases.
There is no noticeable size dependence in both cases.
The obtained value is $z=0$  for the tent and logistic map.  Such low values of $z$ are 
observed only for highly nonlocal updates. For example, $z=0.35$ for Swendsen-Wang
algorithm\cite{swendsen} and  $z=0$ for Wolff algorithm~\cite{PhysRevLett.60.1591}. Even in mean-field limit, $z=0$ for Wolff algorithm~\cite{PhysRevE.54.2351,sokal1991beat}. However,
such low values of $z$ are not obtained for local synchronous updates to our knowledge.

\begin{figure}[ht]
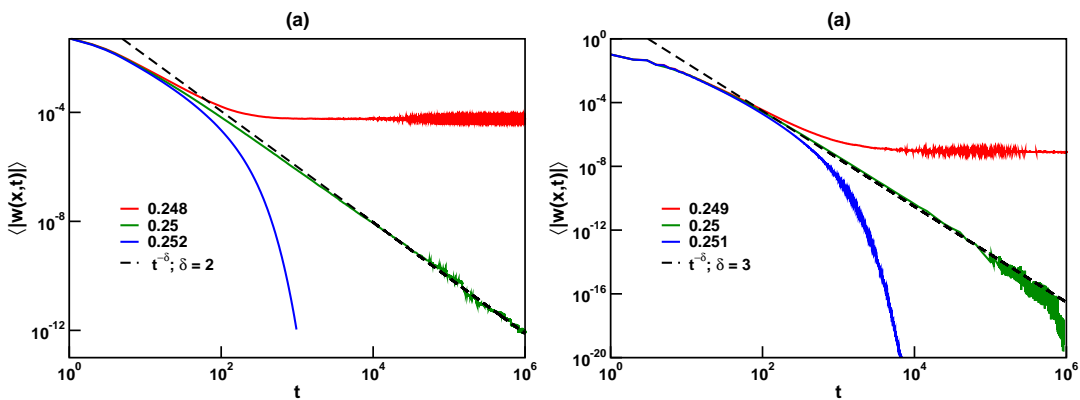

	\centering   
	\includegraphics[width=.45\linewidth]{2dtentop} 
	\includegraphics[width=.45\linewidth]{2dlogop} 
	\caption{(a) Shows the plot of order parameter $\langle|w(x,t)|\rangle$
		with time at $\gamma$, $\gamma<\gamma_{c}$ and$\gamma>\gamma_{c}$
		for coupled 2D tent CML with the disorder in diffusion parameter
		$\epsilon$. At $\gamma_c=0.25$, $\langle|w(x,t)|\rangle$ $\sim t^{-\delta}$ where $\delta=2$.
		(b)Shows the plot of order parameter
		$\langle|w(x,t)|\rangle$ with time at $\gamma$, $\gamma<\gamma_{c}$ and$\gamma>\gamma_{c}$
		for coupled logistic 2D CML with disorder in diffusion parameter $\epsilon$. 
		At $\gamma_c=0.25$, $\langle|w(x,t)|\rangle$ $\sim t^{-\delta}$ where $\delta=3$.} 
	\label{fig5} 
\end{figure}

\begin{figure}[ht]
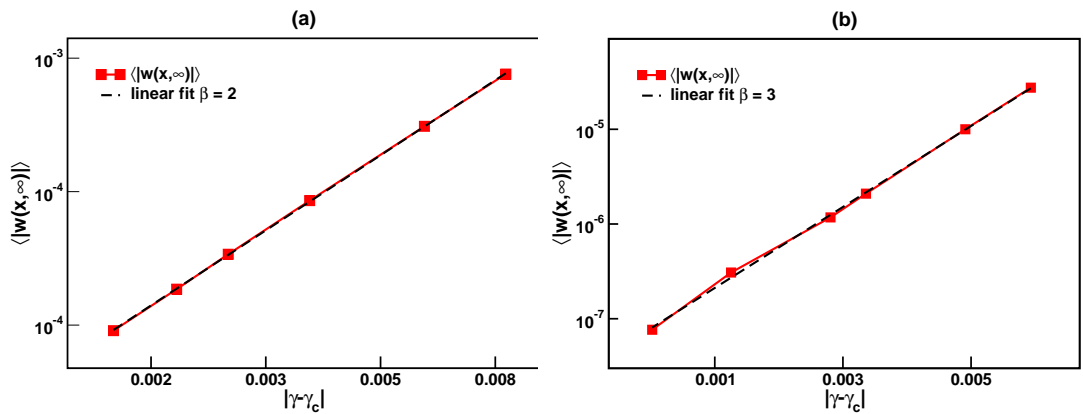

	\centering  
	\includegraphics[width=.45\linewidth]{2dtentbeta} 
	\includegraphics[width=.45\linewidth]{2dlogbeta} 
	\caption{(a)Shows the plot of order parameter $\langle|w(x,\infty)|\rangle$
		for several $\gamma<\gamma_c=0.25$ in the range 0.242-0.248 in the case of tent CML.
		The obtained value of $\beta=2$ in 2D.
		(b)Shows the plot of order parameter $\langle|w(x,\infty)|\rangle$ for several 
		$\gamma<\gamma_c=0.25$ in the range 0.243-0.249 in the case of logistic CML.
		The obtained value of $\beta=3$ in 2D.}
	\label{fig6} 
\end{figure}

\begin{figure}[ht]
	\centering 
	\includegraphics[width=.45\linewidth]{1dtentsize} 
	\includegraphics[width=.45\linewidth]{1dlogsize} 
	\caption{For coupled 1D maps, we show the time evolution of the order
		parameter for various system sizes at the critical point for (a)coupled tent maps.
		(b) coupled logistic maps. There is no noticeable size dependence. } 
	\label{fig11}
\end{figure}

We extend the same model to 2D and global coupling.
The model in the case of 2D tent and logistic maps is as follows. \\
\begin{equation*}
	\begin{pmatrix}
		u_1(i,j,t+1)\\
		u_2(i,j,t+1
	\end{pmatrix}
	=
	\begin{pmatrix}
		1-\gamma & \gamma\\
		\gamma & 1-\gamma
	\end{pmatrix}
	\begin{pmatrix}
		(1-2\epsilon(i,j))f_1(i,j,t)+\frac{\epsilon(i,j)}{2}[G_1(i,j,t)] \\
		(1-2\epsilon(i,j))f_2(i,j,t)+\frac{\epsilon(i,j)}{2}[G_2(i,j,t)]
	\end{pmatrix}
\end{equation*}
with,
\begin{equation*}
	G_1(i,j,t)=f_1(i+1,j,t)+f_1(i-1,j,t)+f_1(i,j+1,t)+f_1(i,j-1,t)
\end{equation*}
\begin{equation*}
	G_2(i,j,t)=f_2(i+1,j,t)+f_2(i-1,j,t)+f_2(i,j+1,t)+f_2(i,j-1,t)
\end{equation*}
Here, $i=1,2,3\dots L$, $j=1,2,3\dots L$ and $t=1,2,3\dots$

Interestingly, we obtain the same exponents even for the 2D case.
We simulate a 2D system of size $L\times L$, where $L=1200$ 
with a quenched disorder in coupling parameter $\epsilon(i,j)=\epsilon^*$. We average over
approximately  $500$ in distinct
disorder configurations and initial conditions and simulate for a long time
$10^6$. We fix $a=2$ for the tent map
and $\mu=4$ in the case of the logistic map.
At $\gamma=\gamma_c=0.25$, we obtain power-law decay of order parameter $\langle|w(x,t)|\rangle$ $\sim t^{-\delta}$, where $\delta=2$ for tent map and $\delta=3$
for the logistic map. For $\gamma>\gamma_c$, 
$\langle|w(x,t)|\rangle \rightarrow 0 $ and 
for $\gamma<\gamma_c$, $\langle|w(x,\infty)|\rangle > 0$ (see Fig.\ref{fig5}).
The characteristic exponent $\beta$ can be obtained from scaling relation  $\langle|w(x,\infty)|\rangle \sim |\gamma-\gamma_c|^\beta$ for $\gamma<\gamma_c$. We obtain
$\beta=2$ for tent map and $\beta=3$ for logistic map
as shown in Fig.\ref{fig6}. The off-critical scaling relation
$\langle|w(x,t)|\rangle$ $\sim t^{\delta \nu_\parallel}t|\gamma-\gamma_c|^{\nu_\parallel}$
is used to obtain critical exponent $\nu_\parallel$.
The obtained value of $\nu_\parallel$=1 for the tent map and logistic map (see Fig.\ref{fig7}).
Again there is no noticeable size dependence and
we can conjecture that $z=0$ in this case as well.

\begin{figure}[ht]
	\centering   
	\includegraphics[width=.45\linewidth]{2dtentoffcr} 
	\includegraphics[width=.45\linewidth]{2dlogoffcr} 
	\caption{(a) shows the off-critical scaling of the order parameter
		$\langle|w(x,t)|\rangle$ for $-0.004\le \gamma-\gamma_c\le 0.008$ in case of 2D tent CML. 
		(b) Shows the off-critical scaling of the order parameter $\langle|w(x,t)|\rangle$
		for $-0.007 \le \gamma-\gamma_c\le 0.007$ in case of 2D logistic CML.
		The best collapse is obtained for $\nu_{\parallel}=1$ in both cases.} 
	\label{fig7} 
\end{figure}

\begin{figure}[ht]
	\centering 
	\includegraphics[width=.45\linewidth]{2dtentsize} 
	\includegraphics[width=.45\linewidth]{2dlogsize} 
	\caption{For coupled 2D maps, we show the time evolution of the order
		parameter for various system sizes at the critical point for (a)coupled tent maps.
		(b) coupled logistic maps. There is no noticeable size dependence for either map.} 
	\label{fig12}
\end{figure}

\begin{figure}[ht]
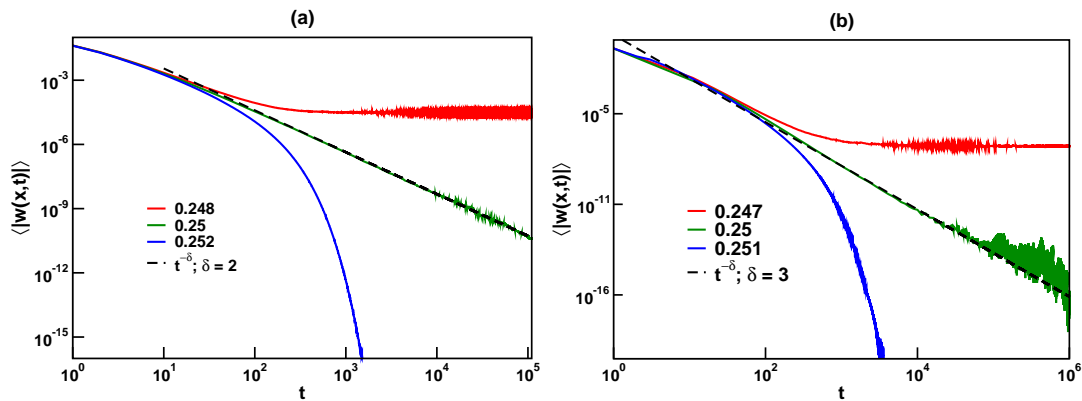

	\centering 
	\includegraphics[width=.45\linewidth]{globtentop} 
	\includegraphics[width=.45\linewidth]{globlogop} 
	\caption{  (a)Shows the decay of  order parameter $\langle|w(x,t)|\rangle$ with time
		at $\gamma<\gamma_c$, $\gamma$ and $\gamma>\gamma_c$. At $\gamma_{c}=0.25$,
		$\langle|w(x,t)|\rangle$ $\sim t^{-\delta}$ where $\delta=2$ for globally coupled 
		1D tent CML with the disorder in diffusion parameter $\epsilon$. 
		(b)Shows the plot of order parameter $\langle|w(x,t)|\rangle$ with time
		at $\gamma<\gamma_c$, $\gamma$ and $\gamma>\gamma_c$. At $\gamma_{c}=0.25$,
		$\langle|w(x,t)|\rangle$ $\sim t^{-\delta}$ where $\delta=3$ in case of globally coupled logistic CML.} 
	\label{fig8} 
\end{figure}

Now we extend the model to global coupling. This is mean-field coupling
and is an effectively infinite dimensional system. Interestingly,
we obtain the same exponents even in this case.
The model for globally coupled maps is as follows.
\begin{equation*}
	\begin{pmatrix}
		u_1(i,t+1)\\
		u_2(i,t+1)
	\end{pmatrix}
	=
	\begin{pmatrix}
		1-\gamma & \gamma \\
		\gamma & 1-\gamma 
	\end{pmatrix}
	\begin{pmatrix}
		(1-2\epsilon(i))f_1(i,t)+\epsilon(i) S_1(i,t)\\
		(1-2\epsilon(i))f_2(i,t)+\epsilon(i) S_2(i,t)
	\end{pmatrix}
\end{equation*}
Here, $S_1(i,t)=\frac{1}{N}\sum_j f_1(i,t)$ and $S_2(i,t)=\frac{1}{N}\sum_j f_2(i,t)$.

\begin{figure}[ht]
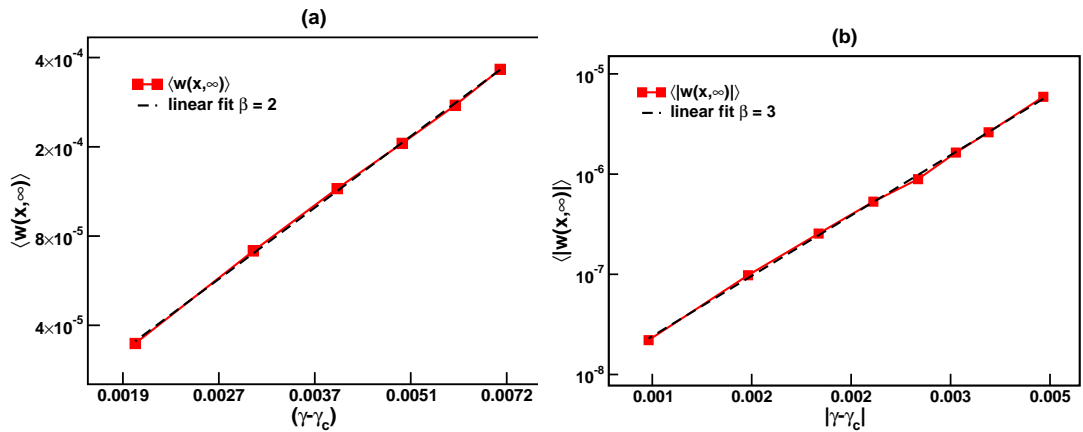

	\centering
	\includegraphics[width=.45\linewidth]{globtentbeta} 
	\includegraphics[width=.45\linewidth]{globlogbeta} 
	\caption{(a) Shows the plot of $\langle|w(x,\infty)|\rangle$ 
		for several $\gamma<\gamma_c=0.25$ in the range 0.244-0.248 in the case of globally
		coupled tent map. The obtained value of $\beta=2$.
		(b) Shows the plot of $\langle|w(x,\infty)|\rangle$ for several $\gamma<\gamma_c=0.25$
		in the range of 0.241-0.249 in the case of a globally coupled logistic map. The obtained value of $\beta=3$.}
	\label{fig9} 
\end{figure}

\begin{figure}[ht]
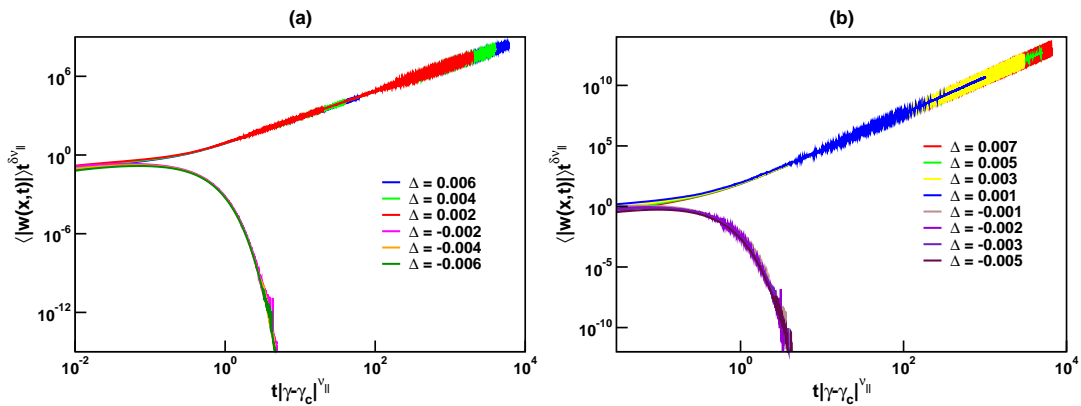

	\centering 
	\includegraphics[width=.45\linewidth]{globtentoffcr} 
	\includegraphics[width=.45\linewidth]{globlogoffcr} 
	\caption{(a) The off-critical scaling of $\langle|w(x,t)|\rangle$
		is shown for $-0.006\le \gamma-\gamma_c\le 0.006$
		for globally coupled tent map.
		(b) The off-critical scaling of $\langle|w(x,t)|\rangle$ is
		shown for $-0.005\le \gamma-\gamma_c\le 0.007$ for
		globally coupled logistic map. The best collapse is
		obtained for $\nu_{\parallel}=1$ in both cases.} 
	\label{fig10}
\end{figure}

We simulate replicas of globally coupled continuous maps of
size $L=5\times 10^5$ for time period $10^6$ and averaged over $500$
distinct disorder and initial condition.
We find that the order parameter synchronization error
$\langle|w(x,t)|\rangle$ undergo synchronization transition in this case
$\langle|w(x,t)|\rangle \sim t^{-\delta}$ with 
$\delta=2$ in case of tent map and $\delta=3$ in case
of logistic map at critical point  $\gamma=\gamma_{c}=0.25$.
For $\gamma>\gamma_c$,$\langle|w(x,t)|\rangle \rightarrow 0 $ and 
for $\gamma<\gamma_c$,$\langle|w(x,\infty)|\rangle > 0$ (see Fig.\ref{fig8}).
$\langle|w(x,\infty)|\rangle \sim (\gamma-\gamma_c)^\beta$ for $\gamma>\gamma_c$ yields
$\beta=2$ in case of tent map and $\beta=3$ in case of logistic map (see Fig.\ref{fig9}).
We obtain $\nu_\parallel=1$ using off critical scaling relation,
$\langle|w(x,t)|\rangle$ $\sim t^{\delta \nu_\parallel} t|\gamma-\gamma_c|^{\nu_\parallel}$
in both cases (see Fig.\ref{fig10}).
There is no size dependence in this case as well and we can conjecture that $z=0$ even in this case.
The critical exponents for both maps in 1D, 2D, and globally coupled lattice are found to be independent of dimension. We conclude that the exponents are superuniversal.

\begin{figure}[ht]
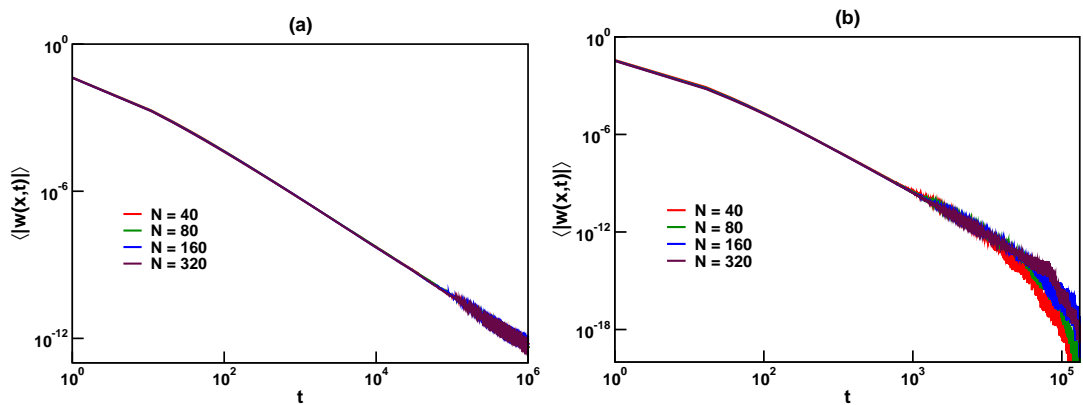

	\centering 
	\includegraphics[width=.45\linewidth]{globtentsize} 
	\includegraphics[width=.45\linewidth]{globlogsize} 
	\caption{(a) System size dependence of globally coupled tent map.
		(b) System size dependence of globally coupled logistic map. } 
	\label{fig13}
\end{figure}

In our numerical investigation $\gamma_c=1/4$ for tent map at $a=2$ and logistic map at
$\mu=4$. This coincidence could be because these maps are topologically conjugate at
the above parameter values and have the same Lyapunov exponent $\log(2)$. However,
if we vary the map parameters, the critical point $\gamma_c$ changes. It 
need not be the same as we change dimensions. We study coupled tent maps for $a=1.5$. 
We find that $\gamma_c=0.1666666$ in 1D and 2D and $\gamma_c=0.142857$ for globally coupled cases. The critical exponent $\delta=2$ in all cases is the same as for $a=2$.
For the logistic map for $\mu=3.8$, we do not find a continuous transition for 
the globally coupled case. However in 1D we find that $\gamma_{c}=0.1781$ and in 2D $\gamma_{c}=0.1783$. The critical exponent is found to be $\delta=1.2$ in both cases.
This exponent is very different from one obtained for $\mu=4$. Thus the critical exponent
depends on the map and its parameter values.

\section{Conclusion}
We study the synchronization of two replicas of coupled map lattice
in the presence of identical quenched disorder in the coupling parameter.
In the absence of quenched disorder, the transition belongs to the MN
universality class for continuous maps. We find that in
the presence of quenched disorder, for coupled tent and logistic map, the system
undergoes a clean second-order phase transition in 1D, 2D as well and
global coupling. The critical exponents are
different from those obtained for the MN universality class.
We find that the values of exponents of 1D, 2D, and global couplings are 
the same. The critical exponent in this case is independent of the dimension
and superuniversal. The order parameter $\langle|w(x,t)|\rangle$ decays as $t^{-\delta}$ at the critical point $\gamma_c=0.25$ in both cases. Here $\delta=2$
for tent map (a=2)  and $\delta=3$ for the logistic map $\mu=4$ in any dimension.
The asymptotic spatiotemporal synchronization error $\langle|w(x,\infty)|\rangle$
grows as $(\gamma-\gamma_c)^\beta$. We find that  $\beta=\delta$ in both cases. The
exponents depend on the details of the map. Thus, the quenched disorder is a relevant perturbation for the replica synchronization of coupled map lattice.  
\section*{Acknowledgment}
PMG thanks DST-SERB for financial assistance (Ref. CRG/2020/003993).
NRS thanks MAHAJYOTI for financial assistance (Ref. MJRF/2022/1002(563)).
\bibliographystyle{iopart-num}
\bibliography{sync}
\end{document}